\documentclass[sigconf,balance=false,authorversion]{acmart}
\AtBeginDocument{%
  \providecommand\BibTeX{{%
    \normalfont B\kern-0.5em{\scshape i\kern-0.25em b}\kern-0.8em\TeX}}}

\copyrightyear{2021} 
\acmYear{2021} 
\setcopyright{acmcopyright}\acmConference[SUI '21]{Symposium on Spatial User Interaction}{November 9--10, 2021}{Virtual Event, USA}
\acmBooktitle{Symposium on Spatial User Interaction (SUI '21), November 9--10, 2021, Virtual Event, USA}
\acmPrice{15.00}
\acmDOI{10.1145/3485279.3485293}
\acmISBN{978-1-4503-9091-0/21/11}

\usepackage{graphicx}
\usepackage{caption}
\usepackage{subcaption}
\usepackage{wrapfig}    
\usepackage{float}
\usepackage{multicol}   
\usepackage{tabularx}        
\usepackage{tabu}

\begin{document}

\title{Adaptic: A Shape Changing Prop with Haptic Retargeting}

\author{J. Felipe Gonzalez}
\email{jf.gonzalez695@uniandes.edu.co}
\orcid{0000-0002-0716-1689}
\affiliation{%
  \institution{Carleton University}
  \city{Ottawa}
  \state{ON}
  \country{Canada}
}

\author{John C. McClelland}
\email{john.mcclelland@carleton.ca}
\affiliation{%
  \institution{Carleton University}
  \city{Ottawa}
  \state{ON}
  \country{Canada}
}

\author{Robert J. Teather}
\email{rob.teather@carleton.ca}
\affiliation{%
  \institution{Carleton University}
  \city{Ottawa}
  \state{ON}
  \country{Canada}
}

\author{Pablo Figueroa}
\email{pfiguero@uniandes.edu.co}
\affiliation{%
  \institution{Universidad de los Andes}
  \city{Bogota}
  \country{Colombia}
}

\author{Audrey Girouard}
\email{audrey.girouard@carleton.ca}
\affiliation{%
  \institution{Carleton University}
  \city{Ottawa}
  \state{ON}
  \country{Canada}
}

\renewcommand{\shortauthors}{J. Felipe Gonzalez, et al.}

\begin{abstract}
  We present Adaptic, a novel "hybrid" active/passive haptic device that can change shape to act as a proxy for a range of virtual objects in VR. We use Adaptic with haptic retargeting to redirect the user’s hand to provide haptic feedback for several virtual objects in arm's reach using only a single prop. To evaluate the effectiveness of Adaptic with haptic retargeting, we conducted a within-subjects experiment employing a docking task to compare Adaptic to non-matching proxy objects (i.e., Styrofoam balls) and matching shape props. In our study, Adaptic sat on a desk in front of the user and changed shapes between grasps, to provide matching tactile feedback for various virtual objects placed in different virtual locations. Results indicate that the illusion was convincing: users felt they were manipulating several virtual objects in different virtual locations with a single Adaptic device. Docking performance (completion time and accuracy) with Adaptic was comparable to props without haptic retargeting.
\end{abstract}



\begin{teaserfigure}
  \centering
     \begin{subfigure}[b]{0.26\columnwidth}
         \centering
         \includegraphics[width=\columnwidth]{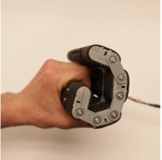}
         \caption{Irregular shape.}
         \label{fig1Letf}
     \end{subfigure}
     \hspace{0.07\columnwidth}
     \begin{subfigure}[b]{0.27\columnwidth}
         \centering
         \includegraphics[width=\columnwidth]{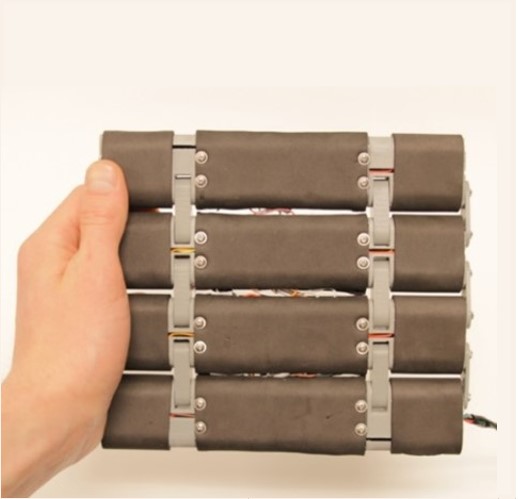}
         \caption{Flat shape.}
         \label{fig1Center}
     \end{subfigure}
     \hspace{0.07\columnwidth}
     \begin{subfigure}[b]{0.26\columnwidth}
         \centering
         \includegraphics[width=\columnwidth]{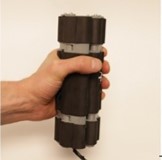}
         \caption{Cylinder shape.}
         \label{fig1Right}
     \end{subfigure}
        \caption{Adaptic can be deformed or actuated to a variety of shapes using its four hinged panels. We used the flat and cylinder shape for the book and flashlight objects in our experiment}
        \label{fig1}
\end{teaserfigure}

\maketitle

\section{Introduction}

Haptics facilitate richer user interactions by adding physical touch feedback to digital interfaces. Haptic feedback is especially valuable in virtual reality (VR). It provides users with a physical reference for virtual objects, improving immersion \cite{slater_place_2009, feick_visuo-haptic_2021} which enhances both presence \cite{aguerreche_reconfigurable_2010,hoffman_physically_1998} and user performance \cite{besancon_mouse_2017,franzluebbers_performance_2018,teather_evaluating_2010}. Most approaches to VR haptics can be characterized as either passive haptic feedback (PHF) or active haptic feedback (AHF). PHF involves the addition of "passive physical objects into virtual environments to physically simulate the virtual objects" \cite{insko_passive_2001}. This often translates in the use of physical proxy objects (i.e., props) that match the shape and position of virtual objects in the scene \cite{besancon_mouse_2017,hu_pneu-multi-tools:_2019,lindeman_towards_1999}, but other recent approaches used tensile bands to generate forces \cite{achibet_elastic-arm_2015}. AHF instead involves using comparatively complex actuated systems \cite{benko_normaltouch_2016,choi_claw_2018,lee_torc_2019, liu_development_2013,massie_phantom_1994,perry_upper-limb_2007} that recreate expected forces from interaction with virtual objects, using, for example, using a controller mounted on a robot arm. In general, PHF is cheap, robust, and supports multiple contact points, but does not generalize well to environments with multiple or different shapes, necessitating a prop for each virtual object. In contrast, AHF generalizes better, supporting virtually any shape, but tend to be prohibitively expensive for most users, are complex and intrusive, limit users' movements, and do not provide multiple contact points.

To leverage the main benefits of both approaches, we developed Adaptic (\autoref{fig1}), a novel shape-changing haptic device for VR. Adaptic is made of four hinged panels and can change shape in real-time to adapt to different haptic shapes, combining the benefits of self-actuated shape-changing interfaces, and relying on visual dominance in VR. Like PHF props, Adaptic supports multiple contact points and provides robust haptic feedback. Similar to AHF devices, the ability to change shape allows Adaptic to generalize to a range of shapes, requiring a single prop rather than many. We consider this an example of "Dynamic Passive Haptic Feedback" (DPHF) \cite{zenner_shifty_2017}, which provides a versatile experience for VR users.

Our goal is to provide realistic haptic feedback in VR experiences requiring manipulation of multiple virtual objects. Adaptic's shape-changing ability presents a solution, as it can simulate a PHF prop for multiple virtual objects while avoiding the need to actually have multiple physical props. However, it cannot match multiple virtual object positions. Thus, we employed haptic retargeting \cite{azmandian_haptic_2016} with Adaptic. Haptic retargeting redirects the user's hand to several virtual object locations within arm's reach, and a single physical prop is used as a proxy for all of them. By using haptic retargeting with Adaptic, users can repeatedly pick up different virtual objects at different virtual locations, while the device changes shape between grasps to provide appropriate tactile feedback for each virtual object being picked up. The users thus think they are picking up different virtual objects and get corresponding tactile cues.

We conducted a study to test our solution compared to the ideal solution of using multiple PHF props. Our goal was to understand how haptic retargeting and the use of different types of props (shape-matching vs. non-shape-matching) affect user experience and performance. Our research questions included:
\begin{itemize}
    \item \textbf{Research question 1 (RQ1):} Can a shape-changing device with haptic retargeting offer a similar experience and performance to using multiple props without retargeting?
    \item \textbf{Research question 2 (RQ2):} How does haptic retargeting impact performance and experience in such scenarios?
    \item \textbf{Research question 3 (RQ3):} Do users prefer shape-matching, or non-shape-matching props?
\end{itemize}

Our contribution is threefold. First, we provide the concept and design behind our novel device Adaptic, which we argue is an example of a DPHF for VR; devices that are minimally actuated just enough to change shape to act as different PHF props. Second, we investigate the potential of using DPHF with haptic retargeting as an alternative to multiple matching props in VR scenarios involving manipulation of multiple objects, a field that has barely been explored. Finally, we investigate haptic retargeting itself, adding to the body of literature on the effectiveness of such techniques.
\begin{figure}
    \centering
    
        \includegraphics[width=0.85\columnwidth]{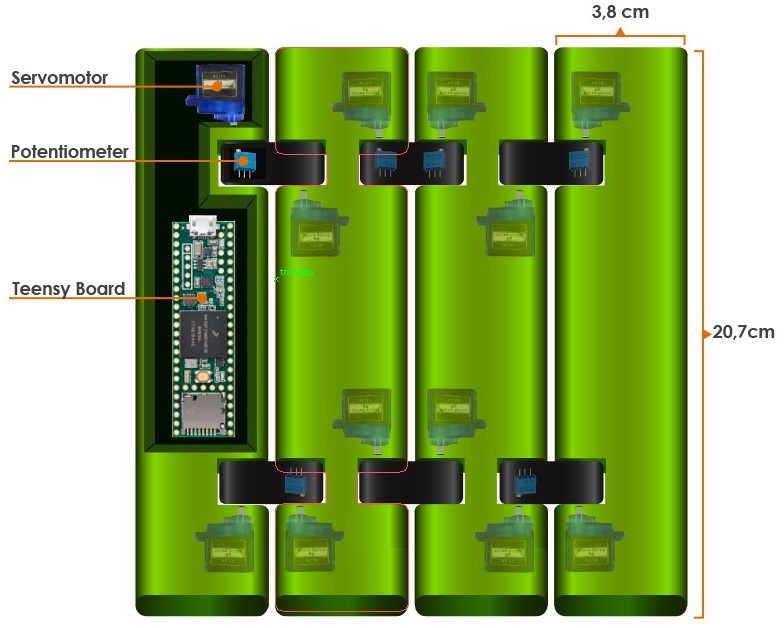}
        \caption{Adaptic hardware configuration layout.} \label{fig3}
  
\end{figure}
\section{Related work}
\subsection{Haptic Feedback in VR}
Research on visual dominance has shown that haptic approximations of virtual objects are sufficient to make users think they are touching the object seen \cite{azmandian_haptic_2016,ban_modifying_2012,dominjon_influence_2005,kwon_effects_2009,simeone_substitutional_2015,zenner_shifty_2017}. Compared to AHF, PHF approaches are simple, inexpensive, and support stronger forces as they are not subject to motor strength. They also support multiple contact points (i.e., full-hand interaction). Some approaches even support a small range of shapes through manual shape-shifting by the user \cite{zhu_haptwist_2019, cheng_iturk_2018, mcclelland_haptobend:_2017}. Haptwist \cite{zhu_haptwist_2019} provides a toolkit to create different interactive props similar to Rubik’s Twist with a design platform that provides several configurations to emulate specific object shapes. iTurk \cite{cheng_iturk_2018} has the user self-actuate PHF props, moving and converting them into different shapes to match objects displayed in VR. Thus, it repurposes one prop to multiple virtual objects through human intervention. PHF props present a simple and practical solution, however, they do not generalize well and necessitate the complexity of switching props; a scene with many unique virtual objects requires a separate prop for each object. 

AHF approaches use motorized devices, such as the Phantom \cite{massie_phantom_1994}, exoskeletons \cite{perry_upper-limb_2007, son_haptic_2018}, wire and pulley systems \cite{liu_development_2013}, or wearable wire and spring systems \cite{fang_wireality_2020, nakao_fingerflex_2020}. Such devices generalize better, as they actuate to limit joint movement and provide force feedback to match the shape of virtual objects. However, AHF devices require expensive, complex, and potentially intrusive systems. They often cannot produce physically robust feedback (i.e., they generate haptic forces which feel “squishy”) \cite{omalley_haptic_2008}. Most systems support only a single contact point, such as a stylus, or a handheld controller, similar to a mouse or 3D tracked VR controllers.
Another approach to AHF devices is to attach hardware to the user’s hands to limit interactions available in VR \cite{liu_development_2013,massie_phantom_1994, son_haptic_2018, nakao_fingerflex_2020}. Son et al.\cite{son_haptic_2018}, Wolverine \cite{choi_wolverine:_2016}, Grabity \cite{choi_grabity:_2017}, Transcalibur \cite{shigeyama_transcalibur_2019}, PuPoP \cite{teng_pupop:_2018}, Wireality \cite{fang_wireality_2020}, and Fingerflex \cite{nakao_fingerflex_2020} are examples of such devices, using rods, springs, voice coils, motors, air pumps, and wires. However, these approaches restrict the type of interactions between objects and hands to just contact, limiting interactions and impacting realism, as it is clear the haptic feedback is not generated by an object.

\begin{figure}
    \centering
    
     \begin{subfigure}[b]{0.45\columnwidth}
         \centering
         \includegraphics[width=\columnwidth]{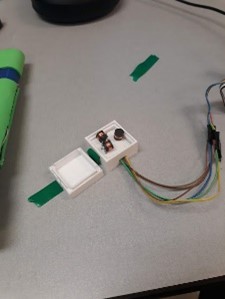}
         \caption{Razer Hydra magnetic coils mounted in a 3D printed box.}
         \label{fig4Letf}
     \end{subfigure}
     \hspace{0.05\columnwidth}
     \begin{subfigure}[b]{0.45\columnwidth}
         \centering
         \includegraphics[width=\columnwidth]{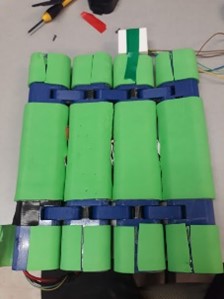}
         \caption{Tracking unit attached to Adaptic}
         \label{fig4Right}
     \end{subfigure}
        \caption{Razer Hydra used for tracking props}
        \label{fig4}

\end{figure}

While our work focuses on approximating object shape, other DPHF projects instead focused on simulating weight and hardness \cite{choi_wolverine:_2016,choi_grabity:_2017, le_goc_zooids:_2016,zhao_robotic_2017,murray_variable_2018, zenner_drag_2019,liu_shapesense_2019,tsai_elastoscillation_2020}. For example, ShapeSense \cite{liu_shapesense_2019}, ElasOscillation \cite{tsai_elastoscillation_2020}, and Drag:on \cite{zenner_drag_2019} change weight distribution and shape dynamically to recreate different mass properties and air resistance feedback for different objects weight. Murray et al. \cite{murray_variable_2018} used a pneumatic controller to simulate different hardness levels with a cylindrical shape. 

Zhao et al. \cite{zhao_robotic_2017} used the Zooid swarm robots \cite{le_goc_zooids:_2016} to construct PHF props out of a set of building blocks. This approach supports a diverse set of props from a limited amount of materials, but the system is complex and limits interactions due to the slow speed of prop construction. Siu et al. \cite{siu_shapeshift:_2018} propose a robot-based solution that moves on a tabletop to give feedback within that space. While the shapes this solution offers is richer than those offered by Adaptic, it does not allow users to grab objects and move them freely. Adaptic extends the concept behind HaptoBend \cite{mcclelland_haptobend:_2017}, a passive shape changing device made of foldable panels that supports a variety of PHF shapes by tracking the device’s shape and rotation in real time. Like Adaptic, HaptoBend applies visual dominance and deformability to VR haptics. However, HaptoBend requires manual intervention to change shape, which limits its dynamic capabilities.

\subsection{Deformable and Shape-Changing Interfaces}
Deformable interfaces allow users to create shapes that complement the context of their use for both fully flexible devices, and those with rigid displays connected by hinges \cite{buschel_foldable3d:_2016,chen_navigation_2008,gomes_paperfold:_2015,hinckley_codex:_2009,ramakers_paddle:_2014}. Most deformable prototypes use a flat, plane-like form factor \cite{gomes_paperfold:_2015,lahey_paperphone:_2011,ramakers_paddle:_2014,steimle_flexpad:_2013}. The use of physical metaphors for input, like bend gestures, also offers potential benefits \cite{holman_paper_2005,khalilbeigi_foldme:_2012,lahey_paperphone:_2011,lo_fabricating_2014,ramakers_paddle:_2014}.

Self-actuated shape-change can communicate notifications with smart phone-like devices \cite{gomes_morephone:_2013,hemmert_shape-changing_2010} and facilitate tangible experiences. Physical interactions through shape-change can add an expressive dimension through force feedback \cite{gomes_morephone:_2013,nakagaki_lineform:_2015,park_trial_2015}, replicate a physical configuration of physical blocks in the virtual realm \cite{follmer_jamming_2012,hawkes_programmable_2010,nakagaki_chainform:_2016,park_trial_2015,roudaut_morphees:_2013}, or facilitate exploration for new interactions \cite{nakagaki_chainform:_2016,ortega_exhi-bit:_2017}.

Device actuation is possible using servo motors \cite{hemmert_shape-changing_2010,lindlbauer_combining_2016,nakagaki_chainform:_2016,rasmussen_balancing_2016,roudaut_morphees:_2013}, shape memory alloys (SMAs) \cite{gomes_morephone:_2013,hawkes_programmable_2010,park_trial_2015,roudaut_morphees:_2013}, particle jamming \cite{follmer_jamming_2012} and linear actuators \cite{ortega_exhi-bit:_2017}. Detailed shape tracking facilitates higher fidelity interactions \cite{gallant_towards_2008,rendl_flexsense_2014,steimle_flexpad:_2013}, but can also capture emotional states \cite{strohmeier_evaluation_2016}, or monitor posture \cite{hermanis_acceleration_2016}. We build on this past work to design a self-actuated shape-changing device.
\subsection{Haptic Retargeting}
Since vision dominates other senses in VR \cite{van_beers_precision_1998,rock_vision_1964}, we can induce perceptual illusions in VR users. The best-known example is redirected walking; the user is made to think they are walking in a straight line, but are in fact walking in arcs and circles via subtle rotations of the virtual environment \cite{razzaque_redirected_2001}. This effectively provides a larger tracking space than physically available \cite{nilsson_15_2018}. 

More recently, Kohli et al. \cite{kohli_redirected_2013} proposed redirected touching. They virtually decoupled the hand from the physically tracked location with minimal impact on user performance. Azmadian et al. \cite{azmandian_haptic_2016} took this idea further and introduced haptic retargeting. The technique involves either redirecting a user’s hand or warping the perceived location of virtual objects relative to a haptic prop. The virtual hand is redirected to approach a given virtual object. Meanwhile, the physical hand touches the same prop repeatedly, convincing the user they are touching a different virtual object in a different location. In fact, they are touching the same physical object and location repeatedly. The sparse haptic proxy employs haptic retargeting, mapping a limited number of physical controls onto a larger number of virtual controls \cite{cheng_sparse_2017}. Other studies have enhanced haptic retargeting to make it less noticeable to users, by shifting the virtual hand representation when the users are not looking at it \cite{lohse_leveraging_2019} or while they are blinking \cite{zenner_blink-suppressed_2021}.

\subsection{Haptic Feedback with Haptic retargeting}

Little research has investigated the combination of haptic retargeting with haptic feedback in specialized scenarios. Feick et al. \cite{feick_visuo-haptic_2021} studied linear translation and stretching across different distances. Moreover, Zenner et al. \cite{zenner_combining_2021} studied the manipulation of different objects using haptic retargeting with a weight-shifting prop. No previous work has explored the idea of manipulating different objects while providing shape fidelity using a shape-shifting DPHF with haptic retargeting.

\section{Adaptic:Design and Development}
 
By combining haptic retargeting with shape-changing devices, a single such device could act as a tactile proxy for multiple differently-shaped objects in a virtual scene, similar to having multiple props. Thus, we designed Adaptic, with two main characteristics:
\begin{itemize}
    \item \textbf{Shape-changing} to transform the device to a specified shape and provide animated haptic feedback; it can actuate from flat to a compact wand-like shape in less than 3 seconds. 
    \item \textbf{Shape-locking} to prevent bending along specified hinges, and to mimic the physical attributes of a virtual object.

\end{itemize}

Due to visual dominance in VR, it is not necessary that our device create exact tactile replicas of virtual objects. We instead approximate a range of shapes using the device, sidestepping the main limitation of using props in VR. Although the number of shapes supported is currently limited by its form factor, this approach supports our current studies on shape shifting using a similar form factor to other devices \cite{mcclelland_haptobend:_2017, zhu_haptwist_2019, cheng_iturk_2018}. Moreover, previous work has shown that approximating haptic shapes is generally sufficient \cite{azmandian_haptic_2016,dominjon_influence_2005,kwon_effects_2009,simeone_substitutional_2015,zenner_shifty_2017}. Overall, Adaptic addresses the underlying issues of complexity, limited interactions, and inadequate haptic feedback.

\subsection{Form Factor} 
To keep our initial prototype simple, we designed it with a similar form factor to HaptoBend \cite{mcclelland_haptobend:_2017} with four foam-covered elliptic flattened cylindrical segments. We built on this previous work as participants generally accepted Haptobend's design as DPHF for multiple shapes, providing a suitable haptic proxy for simple objects such as notebooks, tablets, and flashlights.

We used double-hinged connections allowing 360° rotation of each segment, which allows each segment to fold perfectly flat on top of its neighbour, offering more complex shape options than other hinged devices \cite{mcclelland_haptobend:_2017, gomes_paperfold:_2015}. \autoref{fig1} illustrates various shapes possible with Adaptic, such as flat objects or simple prisms.

\begin{figure}
    \centering
    
             \begin{subfigure}[b]{0.45\columnwidth}
                 \centering
                 \includegraphics[width=\columnwidth]{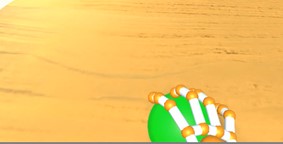}
                 \caption{Participant hand in the home position}
                 \label{fig5_TopLeft}
             \end{subfigure}
             \hfill
             \begin{subfigure}[b]{0.46\columnwidth}
                 \centering
                 \includegraphics[width=\columnwidth]{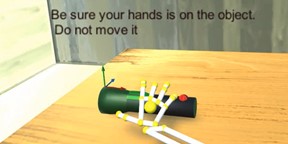}
                 \caption{Part 1: Participant is grasping the target object}
                 \label{fig5_TopRight}
             \end{subfigure}
             \hfill
             \begin{subfigure}[b]{0.46\columnwidth}
                 \centering
                 \includegraphics[width=\columnwidth]{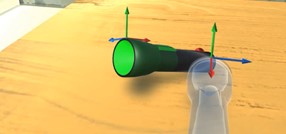}
                 \caption{Part 2: Participant is docking the object with the silhouette at home position}
                 \label{fig5_BottomLeft}
             \end{subfigure}
             \hfill
             \begin{subfigure}[b]{0.46\columnwidth}
                 \centering
                 \includegraphics[width=\columnwidth]{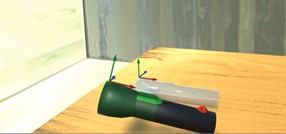}
                 \caption{Part 3: Participant is performing the second docking to the initial position}
                 \label{fig5_BottomRight}
             \end{subfigure}
             \hfill
                \caption{Participant view during the task.}
                \label{fig5}
    
\end{figure}

\begin{figure}
    
        \centering
        \includegraphics[width=0.8\columnwidth]{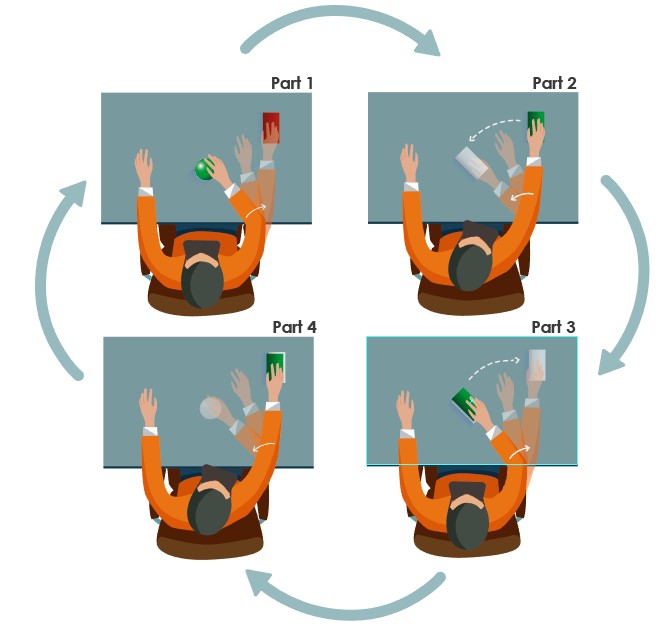}
        \caption{User perception of the task with the right object.} \label{fig7}
 
\end{figure}

\subsection{Hardware}
Adaptic consists of 4 segments articulated with hinged joints (\autoref{fig3}). We designed the segments to be modular and 3D printed them using PLA for assembly with metal screws. The device was actuated by two Mg90S servomotors, powered with a 5DCV supply inside each segment. This gave each servomotor a speed of 60 degrees per 100 ms, and produced a torque of 2.5kg-cm, with a rotation range of 180 degrees. Each joint included a 1KOhm potentiometer (reference 3362p) to measure the joint angle through a simple voltage measurement during shape-change. This allows Adaptic to work as a data-input device while a user changes its shape manually; we did not employ this feature in the current study. We included a Teensy 3.5 board in one segment to control the servomotors, measure joint angles, and manage serial communication with the host computer. Adaptic also tracks its own absolute position and orientation using a Razer Hydra magnetic tracker, discussed further in the Apparatus section.

\subsection{Current Application: Adaptic + Haptic retargeting}

We propose a method for providing tactile feedback in VR scenarios with multiple different objects. Ideally, VR users could manipulate different virtual objects and the system would provide haptic feedback for each, making each virtual object feel like a real one. PHF props offer a potential solution, but present logistical challenges of acquiring and tracking multiple different shaped objects in VR. This problem is exacerbated by each additional virtual object; each requires another prop. Clearly, this solution does not scale. 

Thus, we instead propose the use of Adaptic with haptic retargeting. As detailed above, PHF requires matching both shape and position. Adaptic solves the "shape problem", via its shape-changing ability. We use haptic retargeting to solve the "position problem", of matching virtual object positions to that of the prop. Consider the following example: a user is manipulating object \emph{A} and wants to pick up object \emph{B}, which requires first releasing \emph{A}. At this point, Adaptic changes shape to match \emph{B}. In VR, objects \emph{A} and \emph{B} are in different virtual positions. However, since Adaptic remains in the same physical position, haptic retargeting directs the user's hand towards Adaptic, while the user perceives themselves as reaching towards \emph{B}'s virtual position. Thus, the technique provides haptic feedback for two virtual objects, with different shapes, in different virtual locations using only one DPHF device and haptic retargeting. 

\section{Methodology}

We conducted an experiment to compare Adaptic with haptic retargeting to other haptic approaches (i.e., non-matching props; no retargeting) in both user experience and performance. Our main objective was to determine if users experience convincing haptic feedback with our solution, similar to that provided by props. We assess user performance, to consider overall suitability of our solution. We also sought to understand the influence of haptic retargeting and user reactions to shape-matching vs. non shape-matching props.

\begin{table}[tb]
    \caption{Props used in each haptic approach, Haptic retargeting use (columns) and Shape matching use (rows). }
    \label{lab:ltableConditions}
    \small%
	\centering%
	\setlength\extrarowheight{0.0pt}
  \begin{tabu} to 5\columnwidth {%
	r|%
	*{2}{c}%
	}
    \toprule
    Haptic Approach & Haptic  &  No Haptic  \\ 
    & Retargeting (HR+) &  Retargeting (HR-) \\
    \midrule
    Shape & 1 Adaptic & 1 Book and 1 container \\
    Matching (SM+) & (SM+HR+) &  (SM+HR-)\\ \\
  Non Shape  &  1 Styrofoam ball & 2 Styrofoam balls\\
  Matching (SM-) &(SM-HR+) & (SM-HR-)\\
    \bottomrule
    \end{tabu}
\end{table}

\subsection{Participants}
We recruited 24 participants (aged 20 to 36 years old, mean age of 24 years; 10 men, 14 women). We only recruited right-handed participants, since the retargeting technique depended on this. All had normal or corrected-to-normal vision. Most had little prior VR experience: two used VR systems more than once per week, another two used VR more than once a month, and the rest used VR once a year or less. They received \$10 CAD upon completion.

\subsection{Apparatus}
\subsubsection{Hardware}
We used a PC running Windows 10, with a 4.20 GHz CPU, 32 GB of RAM. We used an Oculus Rift CV1 head-mounted display (HMD) with a resolution of 1080x1200 per eye, and a ~100º field of view. We mounted a Leap Motion onto the front of the Oculus Rift, and used it to track the participant's hand when not gripping a prop, showing a hand model reflecting the user’s hand/finger pose. 

In the experiment, users manipulated a virtual book and flashlight. Depending on the condition, haptic feedback was provided by either Adaptic, matching proxies (i.e., a book and a cylindrical plastic container), or non-matching proxies (i.e., Styrofoam balls). See \autoref{lab:ltableConditions}. All props, including Adaptic, were positioned on the table in front of the participant and tracked with a Razer Hydra magnetic tracker. We used magnetic rather than optical tracking to avoid self-occlusion issues during Adaptic's shape-change. We extracted the magnetic coils from the Hydra controller and mounted them inside a 3D printed box. This box was affixed to each prop and Adaptic with Velcro tape (\autoref{fig4}). The Hydra tracker base was positioned beneath the table, 40 cm from the participant. This provided good tracking quality in the range of motion required for the experiment. To avoid electromagnetic tracking interference, Adaptic's servo motors (which generate their own magnetic field during use) were only active while it changed shape.

When using haptic retargeting, the physical distance between the initial and target positions for the object was 40 cm. The non-retargeting distance was 44.72 cm (40 cm in z-axis ± 20 cm in x-axis). However, in the virtual world, the distance always appeared the same, regardless of haptic approach. 

\subsubsection{Software}
The software was developed in Unity (v2018.3.2f1), and presented a simple scene with the participant seated at a virtual wooden table with a flashlight and book positioned on it. Participants completed a docking task (\autoref{fig5}) alternating between docking the current haptic proxy with a silhouette of either the flashlight or book virtual objects. 

Participants performed 20 trials (10 for each virtual object) for each condition. The task started by pulling the left controller trigger. This made a white “home” sphere appear. The participant then put their virtual hand - tracked by the Leap Motion - in the home sphere. Pulling the trigger again made the home sphere disappear, and the target object silhouette (either the book or flashlight) appear, starting the docking task. The participants' real hand reached towards different positions on the table according to the current condition. Regardless of condition, participants always \textit{perceived} hand motion as going either to the right or the left side of the table to grab the virtual object. In reality, this left/right motion only actually happened in half of the trials, when haptic retargeting was disabled. In the other half of the trials, the hand position was redirected via haptic retargeting (body warping, per Azmandian et al. \cite{azmandian_haptic_2016}) to a prop sitting at the centre of the table, as seen in \autoref{fig8}.

Since using symmetric shape objects could introduce ambiguity in object orientation, we included a small multi-coloured 3 arrow axis (\autoref{fig5_TopRight}, \autoref{fig5_BottomLeft} and \autoref{fig5_BottomRight}), on the virtual objects and silhouettes as a reference for the proper orientation. When a user picked up the object, we did not display the hand model, and instead showed only the object model to avoid problems with tracking an occluded hand with the Leap Motion. 

 In conditions using Adaptic, the software also made it change shape to match the model required: flat for the book (\autoref{fig1Center}) and a cylinder for the flashlight (\autoref{fig1Right}). The software logged the position and orientation of the object and timed each task.

\begin{figure}
    
            \centering
            \begin{subfigure}[b]{0.435\columnwidth}
                 \centering
                 \includegraphics[width=0.72\columnwidth]{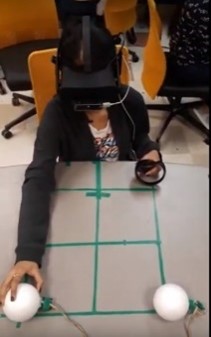}
                 \caption{Non-matching prop, one-to-one mapping.}
                 \label{fig6_Left}
             \end{subfigure}
             \hspace{0.05\columnwidth}
             \begin{subfigure}[b]{0.45\columnwidth}
                 \centering
                 \includegraphics[width=0.95\columnwidth]{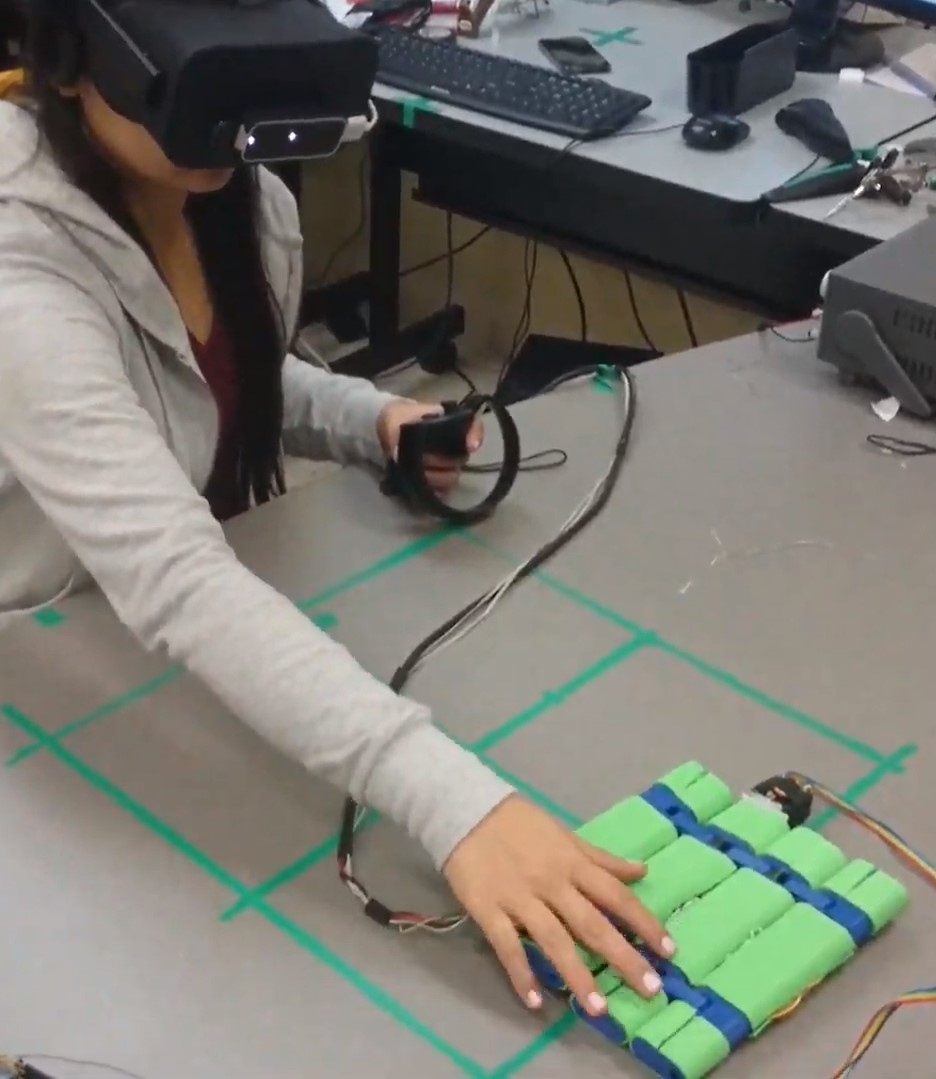}
                 \caption{Shape-matching (via Adaptic) plus retargeting.}
                 \label{fig6_Right}
             \end{subfigure}    
                \caption{Participants performing the docking task.}
                \label{fig6}
    
\end{figure}

\subsection{Procedure}

Upon arriving at our lab, participants provided informed consent prior to participating in the study. We told participants that they would perform docking tasks with different props. Participants were not told about the use of haptic retargeting. 

Next, participants put on the HMD and the experiment began with a brief tutorial. The tutorial always used two non-matching props (i.e., Styrofoam balls) without haptic retargeting. Participants picked up and docked the balls 10 times. Upon completing the tutorial, the participant began with the actual experiment (\autoref{fig6}). 

The experiment task involved docking a virtual book and flashlight with silhouettes. The book was always positioned to the right of the starting position, and the flashlight to the left. The task required moving the virtual objects to the destination and matching their orientation to the silhouette, which was chosen randomly from the five specified angles of the y-axis (-90º, -45º, 0º, 45º, 90º from original orientation). The 0° orientation indicates that the object was aligned with the user’s view direction. The docking task consisted of the following four parts (see \autoref{fig5} and \autoref{fig7}):

\begin{itemize}
    \item \textbf{Part 1—Initial Grab:} The participants moved their hand from the home point to the target object and grabbed it. Their hand disappeared and the object turned green to provide feedback. Participants pulled the left-hand trigger.
    \item \textbf{Part 2—First Docking:} A semitransparent silhouette of the object appeared at the home position, specifying the target position and orientation. The participants moved and docked the object to the specified position and orientation and pulled the trigger when satisfied. 
    \item \textbf{Part 3—Second Docking:} A second semitransparent silhouette then appeared in the object’s original position. The participant again docked the object to this position and orientation and pulled the trigger. 
    \item \textbf{Part 4—Return Home:} Upon releasing the object in its original position, the participant’s hand and the “home” sphere appeared again, and the entire cycle began again. 
\end{itemize}

\begin{figure}
   
            \centering
            \includegraphics[width=0.85\columnwidth]{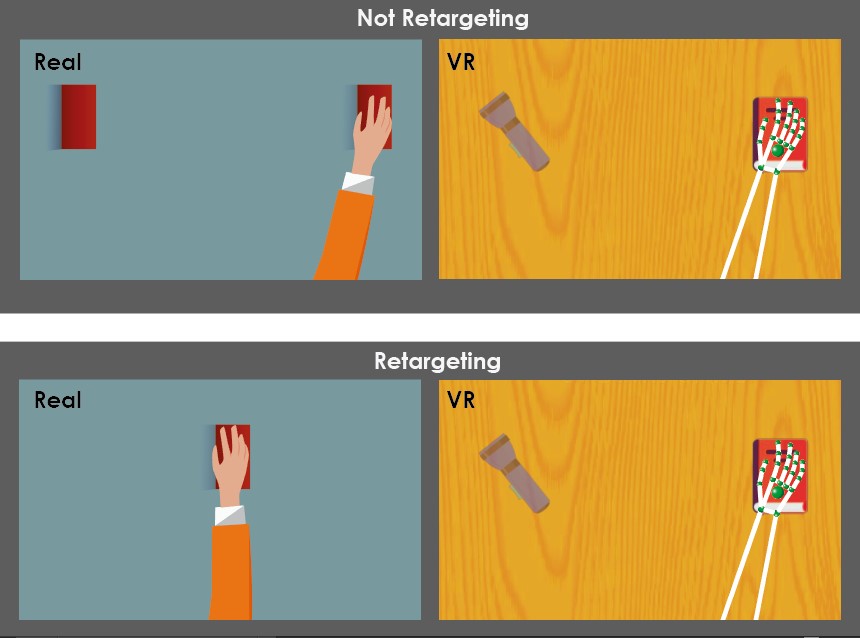}
            \caption{Haptic Retargeting (HR+) and No Haptic Retargeting (HR-) conditions. Left images show real hand motion, right ones virtual hand motion. Top images depict HR- haptic approaches, bottom row ones HR+.} \label{fig8}

\end{figure} 
Participants completed the task as quickly and accurately as possible. After completing all trials with a given haptic approach, participants answered a question about perceived task difficulty in that condition. Upon completing all trials, participants answered a brief questionnaire, including a question intended to help determine if they detected haptic retargeting in half of the trials.

\subsection{Design}

Our experiment employed a 4x2x10 within-subjects design with the following independent variables and levels:
\begin{itemize}
\item Haptic Approach: SM+HR+, SM+HR-, SM-HR+, SM-HR-
\item Object: Book, flashlight
\item Trial: 1, 2, ... 10
\end{itemize}

Haptic approach includes the four combinations of shape-matching props (on/off, represented as SM+ and SM- respectively) and haptic retargeting (on/off, represented as HR+ and HR-, respectively). Each haptic approach thus used a different prop(s), with or without haptic retargeting. The SM+HR+ haptic approach used Adaptic with haptic retargeting, while the SM+HR- haptic approach used matching props (an actual book and cylindrical shaped bottle) with haptic retargeting. Note that although the the props used in the SM+ conditions were not identical between the HR+ and HR- conditions, they still both used shape-matching props, and our goal was to compare the effectiveness of Adaptic to standard props. Both of the SM- haptic approaches used non-matching props, i.e., the Styrofoam balls. The SM-HR+ haptic approach used a single Styrofoam ball with haptic retargeting, while SM-HR- used two Styrofoam balls without haptic retargeting, i.e., one-to-one movement.

Haptic approach order was counterbalanced according to a Latin square. Each trial alternated between the book and the flashlight objects. In total, each participant completed 80 docking trials, for a total of 1920 trials across all 24 participants. 

Dependent variables included included angular error, position error, and completion time (during part 2 "First Docking" and part 3 "Second Docking" of the task). Angular error (in degrees) was the absolute difference between target orientation and actual tracked orientation during docking. Position error (in cm) was the difference between the target object centroid position and manipulated object centroid tracked position in XYZ space. Completion time (in ms) was the time from starting the docking task (i.e., grasping the object) to completion (i.e., releasing the object).

We also used subjective questions to assess participant experience. After completing each haptic approach, participants rated task difficulty with that haptic approach on a 7-point Likert scale (1 = very hard to 7 = very easy) in VR. At the end of the experiment, we conducted a survey to better understand participant experience, with the following questions:

\begin{itemize}
    \item[Q1.] Do you think shape-matching influenced how you experienced the task? (yes/no) 
    \item[Q2.] If yes, which was better? (shape matching, non-matching)
    \item[Q3.] Do you think shape-matching influenced your docking performance? (yes/no)
    \item[Q4.] If yes, which offered better performance? (shape matching, non-matching)
    \item[Q5.] Did you feel a proper correspondence between the real position of your hand and the virtual hand/object? (yes/no)
    \item[Q6.] Rank each prop from 1 Most preferred to 3 Least preferred. (Adaptic/Real props/Styrofoam balls)
\end{itemize}

Note that we explained that non-matching props were the spheres, while all other props were considered "shape matching". We intentionally wrote Q5 without explicitly mentioning haptic retargeting, to avoid bias in participant responses.

\section{Results}

We first present participants’ responses to our subjective questions designed to evaluate user experience. 
We then analyze objective performance results, including task completion time, position error, and angular error for each of parts 2 (first docking) and 3 (second docking) of the task. 

\begin{figure}
\includegraphics[width=\columnwidth]{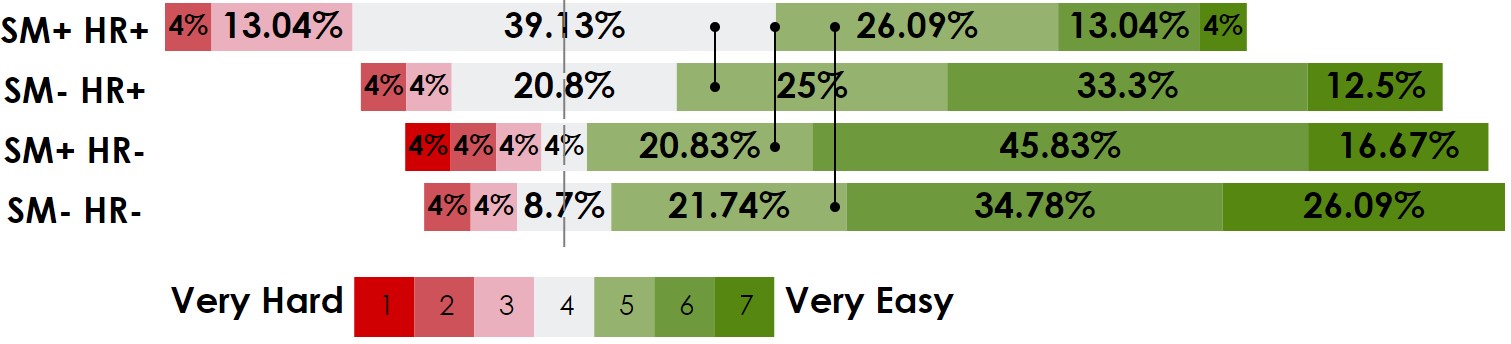}
\caption{Proportion of participant responses on perceived task difficulty for each condition. Black vertical bars show pairwise differences via Conover’s post-hoc test (\(p < .05\)).} \label{fig12}
\end{figure}

\subsection{Subjective results}

We analyzed participants' subjective responses using the Friedman test and Conover’s post-hoc to compare haptic approaches. We also provide basic statistics and insights from the post-questionnaire.

\subsubsection{Subjective Questions Between Conditions} 

According to the Friedman test, there were significant differences in perceived difficulty by haptic approach (\({\chi}_3^2=13.9,p<0.05\)) (\autoref{fig12}). Conover’s post-hoc test at the \(p<.05\) level showed significant differences between SM+HR+ and all other haptic approaches. Participants felt the SM+HR+ was most difficult overall, see \autoref{fig12}.


\subsubsection{Post-Experiment Questionnaire}

Q1 and Q2 tapped into participants’ experience with shape matching. 87.5\% of participants answered that shape matching made a difference in their experience (n=21). Of these, most (85.7\%) felt that shape matching (\(SM\)) improved their experience (n=18). Participants who preferred shape matching indicated that it felt more real and the weight correspondence helped them complete the task. Some participants commented:

\begin{itemize}
    \item “It felt more realistic, and matched my expectations of what to grab.”
    \item “Shape matching improved my accuracy when picking up a prop with the same shape as what is seen.”
    \item “It provided proper visual feedback— what I feel and see match.”
    \item “The affordance of the physical object helped me grip it better.”
\end{itemize}

Q3 and Q4 asked participants about their perceived performance due to shape matching. Again, most participants (79.17\%) believed that prop shape influenced their performance (n=19). Of those, 63.15\% felt shape matching improved their performance (n=12). Several participants commented that they found the weight and grip correspondence made the task more realistic, and due to this familiar sensation, the task became easier. For instance,
\begin{itemize}
    \item “Shape matching helped with weight distribution so I could better match the object.”
    \item “Knowing how to place the object back down depended on the object shape, not the image. My hands knew what I was holding better than my eyes did.”
\end{itemize}

On the other hand, several participants preferred non-matching props (n = 7), indicating that the grip and the lighter weight of the Styrofoam spheres made easier to manipulate the objects and accomplish the task. Some noted:
\begin{itemize}
    \item “The spheres were easier because they were lighter and easier to manipulate.” 
    \item “It was easier to use a sphere—it’s smaller, lighter, and fits in my hand better.” 
\end{itemize}
\begin{figure}
\centering
\includegraphics[width=0.9\columnwidth]{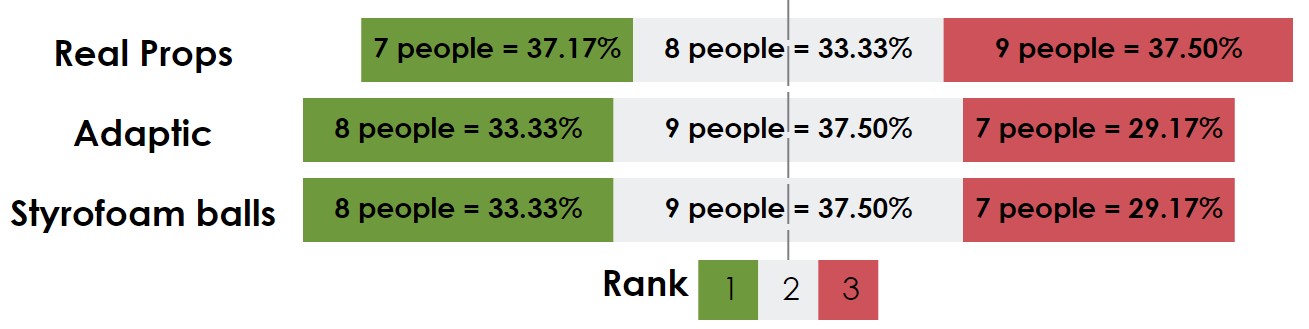}
\caption{Participant prop type preferences. Each prop was ranked from 1 (most preferred) to 3 (least preferred).} \label{fig15}
\end{figure}

Q5 was designed to assess if haptic retargeting had impacted visual and physical perception during the docking task without explicitly priming participants as to the use of haptic retargeting. The question instead was phrased to ask about "good correspondence" between their physical and virtual hand positions. We expected that if haptic retargeting was highly perceptible, participants would likely comment on a poor registration or tracking problems
Of the 24 participants, 20 (83.3\%) answered “yes”, indicating that they felt there was good correspondence between their physical and virtual hands (i.e., they did not notice haptic retargeting). However, of those, 4 mentioned that even though they usually felt the correspondence was good, sometimes there was something "weird" with the tracking, suggesting awareness of the retargeting. For example, 
\begin{itemize}
    \item “I found the system was not 100\% precise. It was 80\% precise to me.” 
    \item “For the last 2 tasks the flashlight seemed to be further to the left on the screen, but in reality it was more to the right.”
\end{itemize}

The 4 participants who answered “no” reported a poor correspondence between the virtual and real hand, noting that it was obvious something was happening. For example, 
\begin{itemize}
    \item “I didn’t feel that I was grabbing the part of the object I was targeting. Sometimes the object in VR didn’t move the same way I was moving the real object.” 
    \item “The objects sometimes felt like they matched up really well, and then other times I missed them when I was grasping at them. Sometimes it was hard to position them because of this.” 
\end{itemize}

At the conclusion of the experiment, we revealed all props and the use of haptic retargeting. We asked about prop preference, asking participants to rank each prop type (matching props, Adaptic, and the Styrofoam balls). See \autoref{fig15}. A Friedman test did not reveal any significant differences in preference.

\subsection{Objective results}

We analyzed completion time, position error, and angular error for two parts of the docking task, part 2 (first docking) and part 3 (second docking), using repeated-measures ANOVA. We first used Mauchly's test (\(\alpha_{\chi}=0.05\)) to verify sphericity. In cases where sphericity was violated, we applied Greenhouse-Geisser corrections (\(\epsilon\)). Finally, we used Fischer’s post-hoc test to detect pairwise significant differences (\(\alpha_{F}=0.05\)). \autoref{figObjectiveData} summarizes our results, including significant pairwise interactions between haptic approach and object.

\subsubsection{Completion Time}

For part 2 completion time, Mauchly's test revealed that sphericity was violated for haptic approach (\(\chi_{5}^2=0.471, \epsilon= 0.693\)), trial (\(\chi_{44}^2=0 , \epsilon= 0.382\)), and the haptic approach x object (\(\chi_{5}^2=0.406 , \epsilon= 0.668\)) and object x trial (\(\chi_{44}^2=0 , \epsilon= 0.385\)) interactions. ANOVA revealed that there were significant main effects on completion time for haptic approach (\(F_{2.08,47.843}=4.234, p=0.019\)), object (\(F_{1,23}=6.079, p=0.022\)) and trial (\(F_{3.438,79.076}=4.301, p=0.005\)). There was also a significant interaction effect between object and trial (\(F_{3.465,79.697}=3.072, p=0.026\)). SM-HR- was significantly faster than the other haptic approaches. We did not find significant differences between the other haptic approaches, including our solution SM+HR+ to the ideal SM+HR-. Docking the book was significantly faster than the flashlight.

\begin{figure*}
\includegraphics[width=\textwidth]{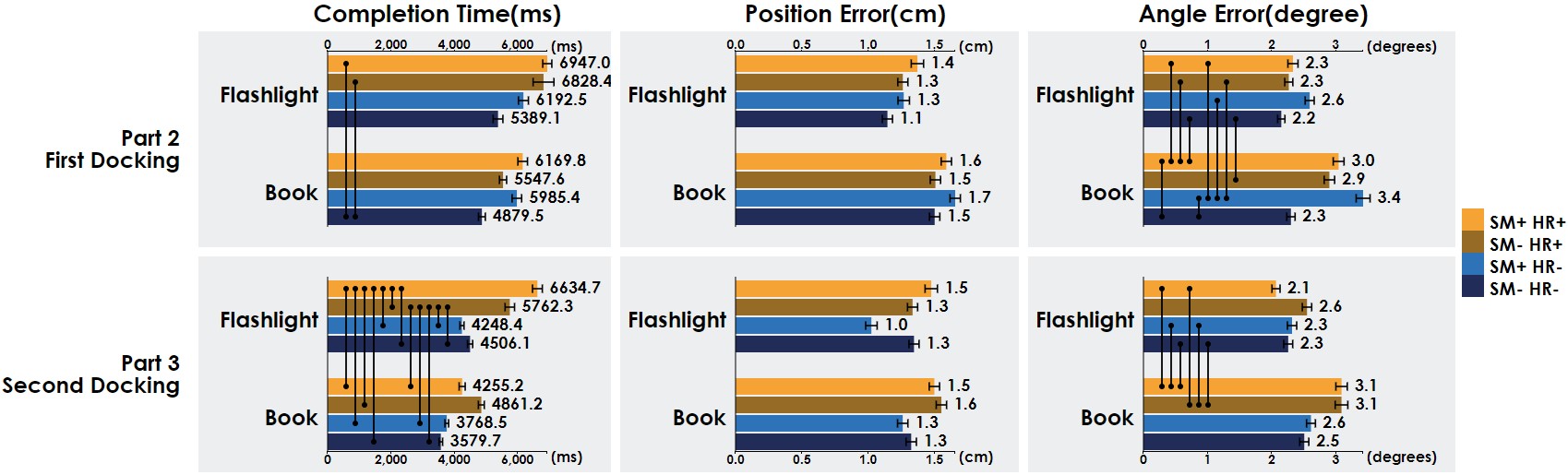}
\caption{Objective data results. Completion time in milliseconds (left), docking position error in cm (center), and docking angular error in degrees (right) for each condition, separated by object, of the task parts 2 (top) and 3 (bottom). Error bars show ±1SE. Black bars show pairwise significant differences via Fischer’s post-hoc test: Completion time, position error and angular error at the \(p < .05\) level.} \label{figObjectiveData}
\end{figure*}

For part 3, sphericity was violated for haptic approach (\(\chi_{5}^2=0.455 , \epsilon= 0.651\)), trial (\(\chi_{44}^2=0.036 , \epsilon= 0.515\)), and the haptic approach x object (\(\chi_{5}^2=0.166 , \epsilon= 0.488\)), and object x trial (\(\chi_{44}^2=0.007 , \epsilon= 0.462\)) interactions. ANOVA revealed significant main effects for haptic approach (\(F_{1.953,44.925}=12.965, p\approx 0.000\)), object (\(F_{1,23}=23.549, p\approx 0.000\)), and trial (\(F_{4.637,106.642}=2.896, p = 0.020\)). The haptic approach x object interaction effect was also significant (\(F_{1.464,33.68}=5.53, p=0.015\)). Both HR+ haptic approaches were significantly slower than the HR- haptic approaches, indicating that haptic retargeting negatively influenced completion time. As with part 2, docking the book was faster than the flashlight. The haptic approach x object interaction indicated that two conditions were significantly slower than all others: Flashlight with SM-HR+ and flashlight with SM+HR+ (\autoref{figObjectiveData} left). 

\subsubsection{Position Error}

For part 2 position error, Mauchly's test indicated sphericity was violated for trial (\(\chi_{44}^2=0.01 , \epsilon= 0.534\)) and object x trial interaction (\(\chi_{44}^2=0.018 , \epsilon= 0.609\)). ANOVA revealed a significant main effect for object (\(F_{1,23}=26.344,p\approx 0.000\)). Neither haptic approach nor trial were significant. Position error was higher with the book than the flashlight, although the differences were small (i.e., less than 0.3cm with std of 0.002cm). This may be because the docking position was slightly elevated from the table, making it harder to dock, especially for awkward-shaped objects. This suggests that object shape influences position error, e.g., due to difficulty in getting the centre lined up.
See \autoref{figObjectiveData} center. 

For part 3, sphericity was violated for haptic approach (\(\chi_{5}^2=0.43 , \epsilon= 0.642\)), trial (\(\chi_{44}^2=0.015 , \epsilon= 0.46\)), and the haptic approach x object interaction (\(\chi_{5}^2=0.498 , \epsilon= 0.666\)). Position error was not significantly affected by any condition. 

\subsubsection{Angular Error}
Angular error scores are seen in \autoref{figObjectiveData} right. For part 2, none of the conditions required Greenhouse-Geisser corrections. ANOVA revealed significant main effects for haptic approach (\(F_{3,69} = 5.337, p=0.002\)) and object (\(F_{1,23}=17.327,p\approx 0.000\)), but not trial. The interaction between haptic approach and trial was also significant (\(F_{27,621}=1.549, p=0.039\)). The SM-HR- haptic approach offered better angular error than the other haptic approaches. There were no significant differences between the other haptic approaches, including between our solution SM+HR+ and the ideal SM+HR-. Docking with the flashlight yielded significantly lower angular error than the book, although the differences are small (i.e., less than 0.7 degrees with std of 0.15 degrees). Note that in part 2, angular error was highest with the book object when using the SM+ haptic approaches (see \autoref{figObjectiveData} right). Shape-matching actually \textit{reduced} angular precision when docking the book.

For part 3, we applied Greenhouse-Geisser correction for haptic approach (\(\chi_{5}^2=0.598 , \epsilon= 0.765\)), trial (\(\chi_{44}^2=0.036 , \epsilon= 0.621\)), and the object x trial interaction (\(\chi_{44}^2=0.042 , \epsilon= 0.592\)). ANOVA revealed a significant main effect for object (\(F_{1,23} = 31.707, p\approx 0.000\)). Error rates were again highest with the book. However, in part 3, the significant differences were when using the HR+ haptic approaches. This may be due to the fact that in part 3, docking positions are more strongly influenced by the retargeting effect. Overall, the angular error in all haptic approaches was small, lower than 3.5°. The book object had higher error rates in general.

\section{Discussion}

\subsection{Adaptic + Haptic Retargeting vs. Props}

For our solution to be viable, it needs to offer comparable user experience to using PHF props for each virtual object (i.e., the haptic approach SM+HR-), without substantially affecting user performance. Subjective preferences towards the various haptic approaches ranked all props (i.e., Adaptic, spheres, book, and container) practically equal, suggesting that Adaptic was about as well-accepted as shape-matching props. However, there was a significant differences in subjective perception of difficulty between our solution, Adaptic with haptic retargeting (i.e., SM+HR+), and the ideal solution of using actual props without haptic retargeting (i.e., SM+HR-). Participants reported finding the actual shape-matching props without haptic retargeting to be easier to use in terms of perceived difficulty than using Adaptic with haptic retargeting. This may be due to better correspondence between the shapes and the virtual objects shown, or perhaps more appropriate weight matching. It may also be directly due to the influence of haptic retargeting itself, causing a slight decoupling between the hand position (as perceived via proprioception) and the virtual hand/object. We note a limitation of our study is the inability to decouple these factors, due to using different shape-matching props in the SM+ conditions. However, neither position error nor angular error were significantly different between those haptic approaches, suggesting that the impact of these factors on performance was relatively small.

Completion time in part 3 of the docking task was more strongly affected by haptic retargeting (i.e., with the SM+HR+ and SM-HR+ haptic approaches), which increased completion time by about 35\%. There are two possible explanations for this. First, the influence of haptic retargeting on the virtual hand position is negligible at the starting point (i.e., the home position) and increases as the hand gets farther from the starting point. We note that of the two docking task components, parts 2 and 3, one involved moving towards the starting point, while the other involved moving away from the starting point. In part 2, the object was moved from the position where haptic retargeting effect was strongest towards where it has a negligible effect. With part 3, this happened in reverse: the task commenced at the starting point where the effect of haptic retargeting was negligible, and moved towards where it was strongest. Thus, participants were more strongly affected when performing movements that increased the influence of haptic retargeting (i.e., moving \textit{away} from the home position in part 3).

Second, the difference in completion time is more pronounced with the flashlight object. The shape itself is unlikely to be the cause, as the highest completion times were seen with both the SM+HR+ and SM-HR+ haptic approaches; the common factor is haptic retargeting. We suspect that it is instead the flashlight's \textit{position} that matters. The flashlight was always located on the left side of the table and all participants were right-handed. As a result, we suspect haptic retargeting's influence was even stronger with the flashlight's left location since such a movement would be farther than reaching to the right side (book) for a right-handed person.

Despite this, SM+HR+ and SM+HR- were comparable in user experience. Even though our solution was perceived as more difficult, overall, approximately 82\% of participant responses here indicated a neutral or better response on task difficulty with SM+HR+ compared to the 92\% in SM+HR- (i.e., very few participants found the task harder when using Adaptic + haptic retargeting). In all part 2 performance results and in all part 3 performance results, except completion time, our solution was comparable. We thus argue that Adaptic is a reasonable candidate for realistic tactile feedback with different shapes when used with haptic retargeting, and might be further improved with a lighter or better form factor.

\subsection{Retargeting and Prop Shape}

The SM+HR+ haptic approach was perceived as significantly harder than the other haptic approaches. This suggests that shape-matching props and haptic retargeting actually made the task harder. This may be explained because the non-matching props were Styrofoam balls that fit easily in the hand, and were lighter than other props. Presumably, this made them easier to manipulate.

Despite participants perceiving greater difficulty with shape matching props, when comparing props directly, they nevertheless expressed preference towards shape-matching props. Interestingly, for quantitative results where the prop shape had a significant effect, and despite participant preference, the SM-HR- haptic approach offered better completion time and angular errors in part 2. In contrast, the SM+ haptic approaches offered worse angular errors in part 2 with the book object. It seems that while performing the task with shape-matching props, participants subconsciously perceived the task as being more difficult. However, the more convincing haptic feedback still influenced their preference towards shape-matching props, perhaps due to improved realism. Participants preferred shape matching props as they improved user experience, in some cases even claiming the performance was better as a result. 

Our results are somewhat inconclusive if participants were aware of haptic retargeting or not. While some might not have noticed, others certainly did, and mentioned there was "something weird" about the real and virtual location correspondence. Moreover, haptic retargeting negatively affected completion time and angular error in part 3. Previous work \cite{azmandian_haptic_2016} reported performance impacts of haptic retargeting; we confirm such effects in more complex object manipulation scenarios. Although this impact reflected statistically significant differences among the haptic approaches, the actual differences where small. We found it interesting that position errors did not increase due to haptic retargeting. As discussed above, the strength of the haptic retargeting effect differed depending on if the manipulated object was to the right or left side of the table, producing larger errors when the distance was larger.

\subsection{Applications}

To determine which style of prop is best in a VR system, one must consider the objectives the system: Is it speed and accuracy? Improved realism? To minimize the number of props? If the objective is to improve presence and enhance realism, then closely emulating physical objects is likely critical \cite{franzluebbers_performance_2018}. For example, in a VR training application, providing haptic proxies that closely mimic tools, including weight and shape, likely better prepares trainees for real circumstances than a default controller. However, in a game or a generic virtual environment with a large number of different-shaped objects, a versatile prop like Adaptic may be useful, despite potential limitations of haptic retargeting.

As such, we argue that Adaptic - or future devices like it - could be applied in scenarios where haptic feedback is required and shape matching fidelity is important. VR games could integrate the connections between emotion and shape-change \cite{lee_emotional_2015,lindlbauer_combining_2016,pedersen_is_2014,strohmeier_evaluation_2016} to bring virtual objects to life. VR narratives could be enhanced by using shape-change to incentivize exploration \cite{nakagaki_chainform:_2016,ortega_exhi-bit:_2017}. Our approach could be used in training applications that require props emulating the physical shape of multiple different objects to provide a "good enough" level of tactile realism. Adaptic can emulate a range of shapes with a single device, with little impact on performance, providing a reasonable alternative to and better user experience than non-matching props. Consider a VR engine repair scenario, which would require many different tools with different shapes and haptic responses. A prop-based solution would be complex. With a single shape-changing device like Adaptic, we could provide a similar and acceptable experience (in terms of presence and performance) at a lower cost and with less tracking complexity. 

\subsection{Limitations}

Our goal was to propose and validate Adaptic with haptic retargeting as an alternative to multiple props. Adaptic was designed to fit multiple shapes to provide better haptic feedback. For this reason, conditions including shape matching props (i.e., SM+HR+ and SM+HR-) used different props in each case (i.e., Adaptic for SM+HR+ vs. a book and container for SM+HR-). Our goal was to compare the ideal solution (matching props without haptic retargeting, i.e., SM+HR-) with our proposed solution (Adaptic, i.e., SM+HR+). Nevertheless, this design decision introduces limitations in our experimental analysis; we cannot decouple the effect of haptic retargeting from the use of different props. While we achieved our objective of testing both alternatives as "complete" solutions, it was not possible to study the effectiveness of Adaptic on its own compared to real props in isolation from the effect of haptic retargeting. This decision was motivated by a desire to keep the experiment size and length relatively short through adding more conditions, but will be revisited in future work. 

Moreover, we did not consider other aspects of haptic feedback such as weight, which can improve haptic experiences and influence user perception of physical \cite{hemmert_shape-changing_2010} and virtual objects \cite{choi_claw_2018,fujinawa_computational_2017,lopes_providing_2017,zenner_shifty_2017}. Indeed, some participants noted the importance of prop weight in task difficulty, which may have explained why Adaptic was deemed slightly more difficult than the other haptic approaches.

Another issue is that despite instructing all participants to prioritize speed and accuracy evenly in the docking tasks, we noticed that some favoured either speed or accuracy. This could yield additional noise in both completion time or position/angle errors, due to the well-known speed-accuracy tradeoff common to such tasks.

We performed an experiment to, among other goals, understand user preferences between shape-matching props and non shape-matching props. We selected the spheres to avoid similarity to the matching props. Results may vary if we had used different non-matching shapes such as cuboids or rectangular prisms. 

Also, we tested haptic retargeting with fixed distances of the real and virtual objects. We thus did not evaluate the impact of distance. A weakness of our experiment design is that it conflated object shape and position, and we can not completely decouple these factors. Moreover, since we only had right-handed participants in the experiment, this may have affected the results when using haptic retargeting for objects positioned on the left, as discussed earlier. We speculate that the inclusion of left-handed participants (had our apparatus permitted it) would likely yield the opposite effect, i.e., a performance cost to objects positioned on the right side of the table, but this is a topic for future study. 

\section{Conclusion}

In this paper, we propose Adaptic, a novel shape-changing device. The device can reconfigure itself to provide a physical proxy for a range of virtual objects in VR applications. Adaptic provides the benefits of shape changing devices and relies on visual dominance in VR to serve as a semi-general-purpose prop. By using haptic retargeting, we can redirect the user’s hand to provide haptic feedback for several virtual objects within the user’s arm reach with only a single device. This provides a better experience than using default shape props (e.g., spheres), and helps solve the scale problem of the ideal scenario of having multiple props for multiple shapes through combining a DPHF device with haptic retargeting.

Overall, our results suggest that Adaptic with haptic retargeting is comparable to the ideal scenario of matching proxies without haptic retargeting, offering only slightly lower performance. Adaptic was ranked as well as other props. We also found that users preferred shape-matching props, despite the perception that they were harder to manipulate and slightly impacted performance in a docking task. Although the participants generally did not mention spatial incoherence between real and virtual objects when using haptic retargeting, we found it had a subtle effect on their performance. Nevertheless, shape-changing devices offer a promising solution for providing PHF for multiple virtual objects. When used with haptic retargeting they provide a versatile and "virtually as good" solution as using multiple PHF props. 

Future work will focus on improving the Adaptic prototype and using it with haptic retargeting in applied scenarios such as training, simulations, and video games. We will also investigate the influence of object position when employing haptic retargeting, and evaluate the impact of revealing a priori the use of haptic retargeting to participants. While our study focuses on one type of shape changing interface in a specific docking task, other devices and tasks may also benefit from the use of haptic retargeting, e.g., HapTwist [67]. Haptic retargeting may be more beneficial in situations where generic shapes are acceptable, or in scenarios where performance is more important than realism.
 
\bibliographystyle{ACM-Reference-Format}
\bibliography{adaptic}

\end{document}